\documentclass[12pt]{article}
\usepackage{amsbsy}
\usepackage{amsmath}
\usepackage{amssymb}
 \usepackage{float}
\usepackage{booktabs}
\usepackage{natbib}
\usepackage{caption}
\usepackage{bm}
\usepackage{mathrsfs}
\usepackage{enumerate}
\usepackage{multirow}
\usepackage{multicol}
\usepackage{array}
\usepackage{color}
\usepackage{epsfig}
\usepackage{graphicx}	
\usepackage{subfigure}
\usepackage{booktabs}
\usepackage{times}
\usepackage[ruled,linesnumbered]{algorithm2e}
\usepackage{booktabs}

\usepackage[titletoc]{appendix}
\usepackage[T1]{fontenc}
\usepackage[utf8]{inputenc}

\usepackage{multirow}

\makeatletter
\def\singlespace{\def\baselinestretch{1}\@normalsize}

\renewcommand{\hat}{\widehat}

\date{\today}  

\begin{document}

\title{Distributed Soft Bayesian Additive Regression Trees}

\author{Ran Hao,Bai Yang
\\ \small Shanghai University of Finance and Economics,Shanghai\ 200433 \\
 \small E-mail:}
\maketitle

\begin{abstract}

Bayesian Additive Regression Trees(BART)  is a Bayesian nonparametric approach  which has been shown to be competitive with the best modern predictive methods such as random forest and Gradient Boosting Decision Tree.The sum of trees structure combined with a Bayesian inferential framework provide a accurate and robust statistic method.BART variant named SBART using randomized decision trees has been developed and show practical benefits compared to BART.
The primary bottleneck of SBART is the speed to compute the sufficient statistics and the publicly avaiable implementation of the SBART algorithm in the R package is very slow.In this paper we show how the SBART algorithm can be modified and computed using single program,multiple data(SPMD) distributed computation with the Message Passing Interface(MPI) library.This approach scales nearly linearly  in the number of processor cores, enabling the practitioner to perform statistical inference on massive datasets. Our approach can also handle datasets too massive to fit on any single data repository.We have made modification to 
this algorithm to make it capable to handle classfication problem which can not be done with the original R package.With data experiments we show the advantage of distributed SBART for classfication problem compared to BART.  
\end{abstract}


Big Data,Distributed Computing, Scalable.

\baselineskip=20pt
\section{Introduction}
\indent
\vspace{-5mm}

Suppose we have a response Y and $p$-dimensional predictor X for n subjects.Consider the regression model
\begin{eqnarray}
    Y_{i}=f_{0}(X_{i})+\varepsilon_{i},i=1, \cdots, n
\end{eqnarray}
where  Gaussian noise $\varepsilon_{i} \sim N\left(0, \sigma^{2}\right) $
and $f_{0}$ is an unknown function of interest.Our goal is to set up a model that can capture relationships $f_{0}$ between X and Y.Often we need more complicated model other than limited linear regression. 

Bayesian Additive Regression Trees(BART)\citep{chipman2010bart} is a non parametric regression model that is often more accurate than other tree-based methods as random forest\citep{breiman2001random},Xgboost\citep{chen2016xgboost}.It loose some stringent parametric assumptions compared to other parametric model. It combines the flexibility of a machine learning algorithm with the formality of likelihood-based inference to create a powerful inferential tool.Another advantage for BART model is its robust performance with respect to various hyperparameter settings and we don't have to waste time on tuning parameter to achieve a perfect fitting.

A problem shared with other tree models is that the resulting estimates of model are step functions,which can introduce error into the model.The BART model archive some degree of smoothing by averaging over the posterior distribution.If the underlying $f_{0}(X)$ is differentiable,we can take advantage of this additional smoothness to get a more accurate model.To introduce smoothness to the model,we can change the decisions made at each node as random rather than deterministic.For example,sample x goes right at branch b of tree $\mathcal{T}$  with probability $\psi(x ; \mathcal{T}, b)=\psi(x ; c_{b}, \tau_{b})=\psi\left(\frac{x_{j}-c_{b}}{\tau_{b}}\right)$  where
$\tau_{b}> 0$ is the bandwidth parameter and $c_{b}$ is the splitting value associated with branch b,$x_{j}$ is the splitting variable.We usually set
\begin{eqnarray}
\psi(x ; c_{b},\tau_{b})=\left(1+e^{-(x-c_{b}) / \tau_{b}}\right)^{-1}
\end{eqnarray}
so that smaller values of x will have higher probability of going left and vice versa.Note that when $\tau\rightarrow 0$ ,the random decision is equal to the deterministic decision of BART.\citet{linero2018bayesianb} refer to trees constructed using the above random decision rule as soft trees and call this BART variant as SBART.They also showed the substantial theoretical and practical benefits for SBART.The primary drawback is that it needs to compute every node's $\psi(x ; \mathcal{T}, b)$ for every sample x,rather than just a single leaf node in BART model.They mentioned that the SBART is the slowest package among the competitors .Actually the slowness impede the application of the SBART.In this paper,we try to accelerate SBART with distributed computing and compare it with a  embarrassingly parallel (i.e. no communication overhead) algorithm.Another reason we choose distributed computing is that we keep data at different location.For some reasons the data are not permitted to send out.Only summary statistics can be derived. 

Some work have been done to consider BART in distributed situation.\citet{pratola2014parallel}  showed how the BART algorithm is modified and how to compute using single program, multiple data (SPMD) parallel computation implemented using the Message Passing Interface (MPI) library. \citet{2017Parallel}  explore a parallel implementation of BART using Apache Spark framework and mentioned that from the speed perspective the MPI version is faster.This paper is mainly based on the work of  \citep{pratola2014parallel} to develop distributed SBART.We also optimize the SBART algorithm and speed up the calculation and at the same time reduce the data size need to exchange.We also expand the SBART model to classification problem.

The rest of the paper proceeds as follows. Section 2 reviews some important details of the MCMC algorithm for fitting BART and SBART so that readers may understand how the optimization can be carried out. Section 3 explains how we have implemented an efficient and distributed version of the SBART algorithm.Section 4 compares the actual times needed to run serial SBART   and the distributed SBART implementation in data experiments.We also show advantage for SBART in classification problem. We summarize our findings in Section 5.

\section{The BART And SBART Algorithm}
We first review those aspects of the BART and SBART methodology  for the understanding of this paper.

\subsection{ The Sum of Trees Model}

For a p-dimensional vector of predictors $X_{i}$ and a response $Y_{i} (1\leq i \leq n)$ ,the BART model posits
\begin{eqnarray}
    Y_{i}=f(X_{i})+\varepsilon_{i},  \varepsilon_{i} \sim N\left(0, \sigma^{2}\right) ,i=1, \cdots, n
\end{eqnarray}
To estimate $f(X)$, a sum of regression trees is specified as

\begin{eqnarray}
    \hat{f}(X_{i})=\sum_{j=1}^{m} g\left(X_{i} ; T_{j}, M_{j}\right)
\end{eqnarray}
$T_{j}$ is the $j^{th}$ binary tree structure and  $M_{j}=\left\{\mu_{1 j}, \ldots, \mu_{b_{j}}\right\}$is the terminal node parameters associated with $T_{j}$ .$T_{j}$ contains information of which bivariate to split on ,the cutoff value ,as well as the internal node's location.
$m$ denote the number of trees which is usually fixed at 200 or 50.

 \subsection{Prior }
 The prior distribution for BART model is $P\left(T_{1}, M_{1}, \ldots, T_{m}, M_{m}, \sigma\right)$.
 Here we assume that
 $\left\{\left(T_{1}, M_{1}\right), \ldots,\left(T_{m}, M_{m}\right)\right\}$ are independent with $\sigma$ ,and the $i^{th}$ tree $\left(T_{i}, M_{i}\right)$ is independent with the $j^{th}$ tree $\left(T_{j}, M_{j}\right)$ when $i \ne j$,so we have
  \begin{eqnarray}
  \label{equ:s1}
 \begin{aligned}
P\left(T_{1}, M_{1}, \ldots, T_{m}, M_{m}, \sigma\right) &=P\left(T_{1}, M_{1}, \ldots, T_{m}, M_{m}\right) P(\sigma) \\
& =\left[\prod_{j}^{m} P\left(T_{j}, M_{j}\right)\right] P(\sigma) \\
&=\left[\prod_{j}^{m} P\left(M_{j} \mid T_{j}\right) P\left(T_{j}\right)\right] P(\sigma) \\
&=\left[\prod_{j}^{m}\left\{\prod_{k}^{b_{j}} P\left(\mu_{k j} \mid T_{j}\right)\right\} P\left(T_{j}\right)\right] P(\sigma)
\end{aligned}
  \end{eqnarray}
So we need to specify the prior for $\mu_{k j} \mid T_{j}, \sigma,$ and  $T_{j}$.
\\
For the convenience of computation, we use the conjugate normal distribution $N\left(\mu_{\mu}, \sigma_{\mu}^{2}\right)$ as  the prior for $\mu_{i j} \mid T_{j}$,$(\mu_{\mu}$,$\sigma_{\mu})$can be derived through computation.
 \\
 The prior for $T_{j}$ is specified by three aspects:
 \begin{itemize}
  \item [1)]
  The probability for a node at depth $d$ to split ,given by $\frac{\alpha}{(1+d)^{\beta}}$ .We can confine the depth of each tree by control the splitting probability so that we can avoid overfitting.Usually $\alpha$ is set to 0.95 and $\beta$ is set to 2.
  \item [2)]
  The distribution on the splitting variable assignments at each interior node, default as  uniform distribution.\citet{rockova2017posterior,linero2018bayesianb} introduced Dirichlet distribution for high dimension variable selection scenario.
  \item [3)]
  The distribution for splitting value assignment,default as uniform distribution.
\end{itemize}

We use a conjugate prior, here the inverse chi-square distribution for prior of $\sigma$,$\sigma^{2} \sim v \lambda / \chi_{v}^{2}$,the two parameters $\lambda$,$v$ can be roughly derived by calculation.
\subsection{Posterior Distribution}

With the settings of priors $(\ref{equ:s1})$,the posterior distribution can be obtained by
\begin{eqnarray}
\label{equ:s2}
\begin{aligned}
P\left[\left(T_{1}, M_{1}\right), \ldots,\left(T_{m}, M_{m}\right), \sigma \mid Y\right] \propto & P\left(Y \mid\left(T_{1}, M_{1}\right), \ldots,\left(T_{m}, M_{m}\right), \sigma\right) \\
& \times P\left(\left(T_{1}, M_{1}\right), \ldots,\left(T_{m}, M_{m}\right), \sigma\right)
\end{aligned}
\end{eqnarray}
which can be obtained by  Gibbs sampling.We need to calculate m successive
\begin{eqnarray}
\label{equ:s3}
\begin{aligned}
P\left[\left(T_{j}, M_{j}\right) \mid T_{(j)}, M_{(j)}, Y, \sigma\right]
\end{aligned}
\end{eqnarray}
 where $T_{(j)}$ and $M_{(j)}$ consist of
all the trees information and parameters except the $j^{th}$ tree.Then $P\left[ \sigma  \mid  \left(T_{1}, M_{1}\right), \ldots,\left(T_{m}, M_{m}\right), Y \right]$ can be obtained from sample from  inverse gamma distribution with explicit expression.
\\

How to draw from $P\left[\left(T_{j}, M_{j}\right) \mid T_{(j)}, M_{(j)}, Y, \sigma\right]$ ? Note that $T_{j}$, $M_{j}$ depends on  $ T_{(j)}, M_{(j)}$ through
$R_{j}=Y-\sum_{w \neq j} g\left(X, T_{w}, M_{w}\right)$
 ,it's equivalent to draw posterior from a single tree of
\begin{eqnarray}
\label{equ:s4}
P\left[\left(T_{j}, M_{j}\right) \mid R_{j}, \sigma\right]
\end{eqnarray}
We then split $(\ref{equ:s4})$ in two steps.First we draw from $P\left(T_{j} \mid R_{j}, \sigma\right)$,then draw posterior from $P\left( M_{j} \mid T_{j},  R_{j}, \sigma\right)$.

In the first step,we have
  \begin{eqnarray} \label{equ:s5}
  P\left(T_{j} \mid R_{j}, \sigma\right) \propto P\left(T_{j}\right) \int P\left(R_{j} \mid M_{j}, T_{j}, \sigma\right) P\left(M_{j} \mid T_{j}, \sigma\right) d M_{j}
  \end{eqnarray}
 ,we call $\int P\left(R_{j} \mid M_{j}, T_{j}, \sigma\right) P\left(M_{j} \mid T_{j}, \sigma\right) d M_{j}=P\left(R_{j} \mid T_{j}, \sigma\right)$ as marginal likelyhood.Because  conjugate Normal prior is employed for $ M_{j}$,we can get an explicit expression of the marginal likelihood.We generate a candidate tree $T_{j}^{*}$ from the previous tree structure $T_{j}$ using  MH algorithm.
We accept the new tree structure $T_{j}^{*}$ with probability

  \begin{eqnarray} \label{equ:s6}
  \alpha\left(T_{j}, T_{j}^{*}\right)=\min \left\{1, \frac{q\left(T_{j}^{*}, T_{j}\right)}{q\left(T_{j}, T_{j}^{*}\right)} \frac{P\left(R_{j} \mid X, T_{j}^{*}\right)}{P\left(R_{j} \mid X, T_{j}\right)} \frac{P\left(T_{j}^{*}\right)}{P\left(T_{j}\right)}\right\}.
  \end{eqnarray}
   $q\left(T_{j}, T_{j}^{*}\right)$ is  the probability for the previous tree $T_{j}$ moves to the new tree $T_{j}^{*}$.
The candidate tree $T_{j}^{*}$ is proposed using four type of moves:
 \begin{itemize}
  \item [1)]
Grow,splitting a current leaf into two new leaves.
  \item [2)]
Prune,collapsing adjacent leaves back into a single leaf.
  \item [3)]
Swap,swapping the decision rules assigned to two connected interior nodes.
  \item [4)]
Change,reassigning a decision rule attached to an interior node.
\end{itemize}
Note that in SBART package,the swap move is not adopted.

Once we have finished sample from  $P\left(T_{j} \mid R_{j}, \sigma\right)$,we can sample the $k^{th}$ tree the $j^{th}$ leaf parameter $\mu_{k j}$ from a explicit normal distribution.With all the process described above we can iteratively sample from the posterior distribution and obtain  valid estimation by merging sampled results.

\subsection{ The SBART Algorithm}

With the definition of the  logistic gating function $\psi(x)$,the probability of going to leaf $\ell$  is
  \begin{eqnarray}
\phi(x ; \mathcal{T}, \ell)=\prod_{b \in A(\ell)} \psi(x ; \mathcal{T}, b)^{1-R_{b}}(1-\psi(x ; \mathcal{T}, b))^{R_{b}}
  \end{eqnarray}
  where $A(\ell)$ is the set of ancestor nodes of leaf $\ell$ and $R_{b}=1$ if the path to $\ell$ goes right at $b$.Here we denote $\phi_{i}$ as the probability vector for the ith sample $x_{i}$ to go to each leaf of the tree.

  We get the marginal likelihood with explicit expression

  \begin{eqnarray}
P\left(R_{j} \mid T_{j}, \sigma,\sigma_{\mu}\right)=\frac{|2 \pi \Omega|^{1 / 2}}{\left(2 \pi \sigma^{2}\right)^{n / 2}\left|2 \pi \sigma_{\mu}^{2} \mathrm{I}\right|^{1 / 2}} \exp \left(-\frac{\|R_{j}\|^{2}}{2 \sigma^{2}}+\frac{1}{2} \widehat{\mu}^{\top} \Omega^{-1} \widehat{\mu}\right),
  \end{eqnarray}
where
  \begin{eqnarray}
\Omega=\left(\frac{\sigma_{\mu}^{2}}{T} \mathrm{I}+\Lambda\right)^{-1}, \quad \Lambda=\sum_{i=1}^{n} \phi_{i} \phi_{i}^{\top} / \sigma^{2}, \quad \widehat{\mu}=\Omega \sum_{i=1}^{n} R_{i} \phi_{i} / \sigma^{2}
  \end{eqnarray}

So the sufficient statistics for marginal likelihood is $(\sum_{i=1}^{n} \phi_{i} \phi_{i}^{\top},\sum_{i=1}^{n} R_{i} \phi_{i}$) which plays important roles in the distributed system.

Besides the randomized decision rule,SBART use a sparsity-inducing Dirichlet  as prior distribution for splitting variables 
so it can adapted to high dimensional scenario for variable selection.

\section{Distributed SBART}

In this section we show how we optimize the original SBART and turn it into a distributed computation.
\subsection{Distributed Computing}

Distributed computing has attracted increasing attention nowadays along with the real-world datasets  become larger and increasingly complex.Some frameworks are developed to handle these situations such as Hadoop,Apache Spark  and Message Passing Interface.

Apache Spark is a cluster computing framework that executes the applications much faster than Hadoop.
One of it's advantage is that it can handle situation when some node in the cluster fail.
\citet{2017Parallel}  explored a parallel implementation of BART using Apache Spark framework and mentioned that
the parallel BART package with MPI support is  superior as it minimizes communication costs and the source code is writing in C++ , a faster programming language.

\citet{pratola2014parallel} built parallel BART  with Message Passing Interface support.Their version of BART only allows for growing and pruning steps in the MCMC sampler with a little efficiency lost. They argue that these two steps are sufficient as the trees are small and therefore easily explored.When we use MPI type of distributed computing,the data is partitioned among workers and each worker can work on its own data and communicate with each other by sending messages.We construct the distributed SBART with the help of MPI.
Despite the difference generated from random numbers,the distributed version of SBART can get the same result from serial version of SBART.

\subsection{Distributed SBART}

We can find in the algorithm of SBART that we need to calculate the sufficient statistics 5 times in one tree update.To sample tree structure $\mathcal{T}_{t}$,we need to calculate it twice.To sample bandwidth $\tau_{t}$,we also need twice calculation.And to obtain the tree node parameter estimation $\mathcal{M}_{t}$,another calculation is needed.In fact we can reduce 2 times of sufficient statistics calculation.That is,after we sample tree structure $\mathcal{T}_{t}$,we record the latest result of marginal likelihood and the corresponding tree node parameter estimation $\mathcal{M}_{t1}$ .When we  sample $\tau_{t}$,we only need to calculate the sufficient statistics for the new proposed $\tau_{t}$ and the corresponding tree node parameter estimation $\mathcal{M}_{t2}$ .Then use the two group of sufficient statistics to judge whether or not to update the bandwidth.Based on the judgement,we can update the tree with the corresponding parameter estimation $\mathcal{M}_{t1}$ or $\mathcal{M}_{t2}$.

\begin{figure}
	
	\begin{center}
		\includegraphics[width=0.6\textwidth]{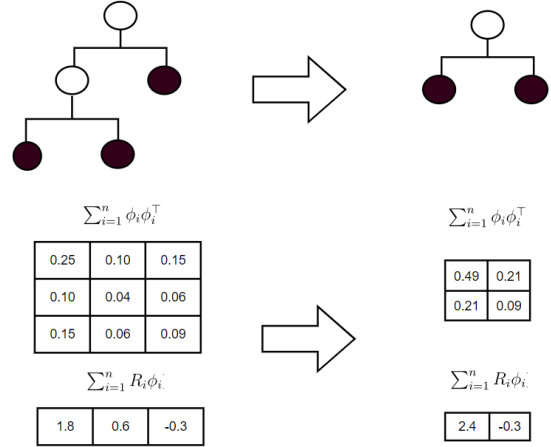}
	\end{center}
	\caption  {  \small{\it{Calculation the new sufficient statistics from previously calculated  sufficient statistics when prune step is adapted.}}}
	\label{FIG1}
\end{figure}

When we sample tree structure $\mathcal{T}_{t}$,the two groups of sufficient statistics are the same for part of the matrix or vector.We should pay attention to part been changed and ignore the unchanged part.By this means,we can accelerate the algorithm and also reduce information size need to transfer.For example,we sample new tree structure $\mathcal{T}_{t}$ with a move of prune.First we get the  sufficient statistics
$(\sum_{i=1}^{n} \phi_{i} \phi_{i}^{\top},\sum_{i=1}^{n} R_{i} \phi_{i}$)  for the old tree structure,we don't have to calculate all part of the sufficient statistics for the new tree ,we can derive it directly from the old sufficient statistics.In figure $(\ref{FIG1})$ we show an example of prune in which case we can directly obtain the sufficient statistics by merging the information of the two nodes to be pruned together.With change step or grow step,we can only focus on the two nodes involved so we can save some time to calculate and reduce the information need to be transfer between workers and master.

Another try we failed is that we try reduce the process time by saving the  $\phi$ for latter use.In big data scenario,$\phi$ become a very large matrix.On the contrary,it need much resources and is slower against our expectation.

We try stop updating the $\tau_{t}$ or attach  $\tau_{t}$ with each node and find the estimated accuracy become worse.It's not beneficial enough to save a little time in exchange of the accuracy.

With all the effort we try to improve SBART in the running time.In Table $\ref{tb1}$ we show the performance difference between  serial SBART and optimized distributed SBART with 2 workers.We can assume that distributed SBART with 2 workers will cost about half of the time optimized SBART with 1 worker need.So we can find that the optimization version is 1.5 times faster for small size up to 2 times faster for large sample size.

\begin{table}
	\begin{center}
		\caption{Performance of serial SBART versus the distributed SBART with 2 workers for moderately sized datasets. Both were run on a simulated dataset with 10 covariates using 1,000 MCMC iterations with the first 500 discarded as burn-in.}
		\label{tb1}
		\begin{tabular}{lcc}
			\hline\hline
			& Serial SBART & Optimized SBART with 2 workers \\ \hline
			1000  & 41.9         & 15.9            \\
			2000  & 78.5         & 28.8            \\
			4000  & 157.5        & 51.0            \\
			8000  & 307.5        & 101.0           \\
			16000 & 843.7        & 207.0           \\ \hline\hline
		\end{tabular}
	\end{center}
\end{table}

Then we outline our distributed algorithm here.Given K workers numbered $0,1,2, \ldots, K-1$,we split the data $(Y,X)$ into $K$ approximately equally-sized portions.The $i^{th}$ data portion $(Y_{i},X_{i})$ resides on worker i .The tree structures $\left(\left(T_{1}, M_{1}\right),\left(T_{2}, M_{2}\right), \ldots,\left(T_{m}, M_{m}\right)\right)$ are kept in every worker.The algorithm follow the  master-slave arrangement,where the numbered $0$ core is regarded as a master and the other are slaves.

There is one difference between our algorithm and the work of \cite{pratola2014parallel}.They split the data into $K-1$ portions and leave the  master worker with no data.They only assign the MCMC sampler job to the master core.After the MCMC sampler job has been done,the master core will idle all the time and it's kind of waste computing resources.So we assign one proportion of data to the master core and make full use of the computing resources.It's helpful when the number of workers is not big.Even when the number of workers is large,we can reduce the data size in the master worker so it can pay more attention to coordinating and communicating with slaves workers.

Considering the draw $(\sigma \mid T_{1}, \ldots T_{m}, M_{1}, \ldots, M_{m}, y) $ as a simple example of message passing, 
we need compute the sufficient statistic $\sum_{i=1}^{n} \epsilon_{i}^{2},$ where $n$ is the total number of observations
and $\epsilon_{i}=y_{i}-\sum_{j=1}^{m} g\left(x_{i} ; T_{j}, M_{j}\right)$.So each worker compute the $\epsilon_{i}^{2}$ in its data portion,sum up them and send the results to the master worker.After the master collect all the result and can draw out a new $\sigma$.Then the master will distribute the new $\sigma$ to all the slaves to keep the synchronization of information at different worker.In distributed computing we must know how to decompose sufficient statistics into corresponding job that each worker can carry out with its own data proportion.Not all the statistics need to be processed by this way.For example when sampling the splitting variable $s$,the updated $s$ only need to keep in the master core and has no need to broadcast it out to slave workers.
\begin{figure}
	\begin{center}
		\includegraphics[scale=0.4]{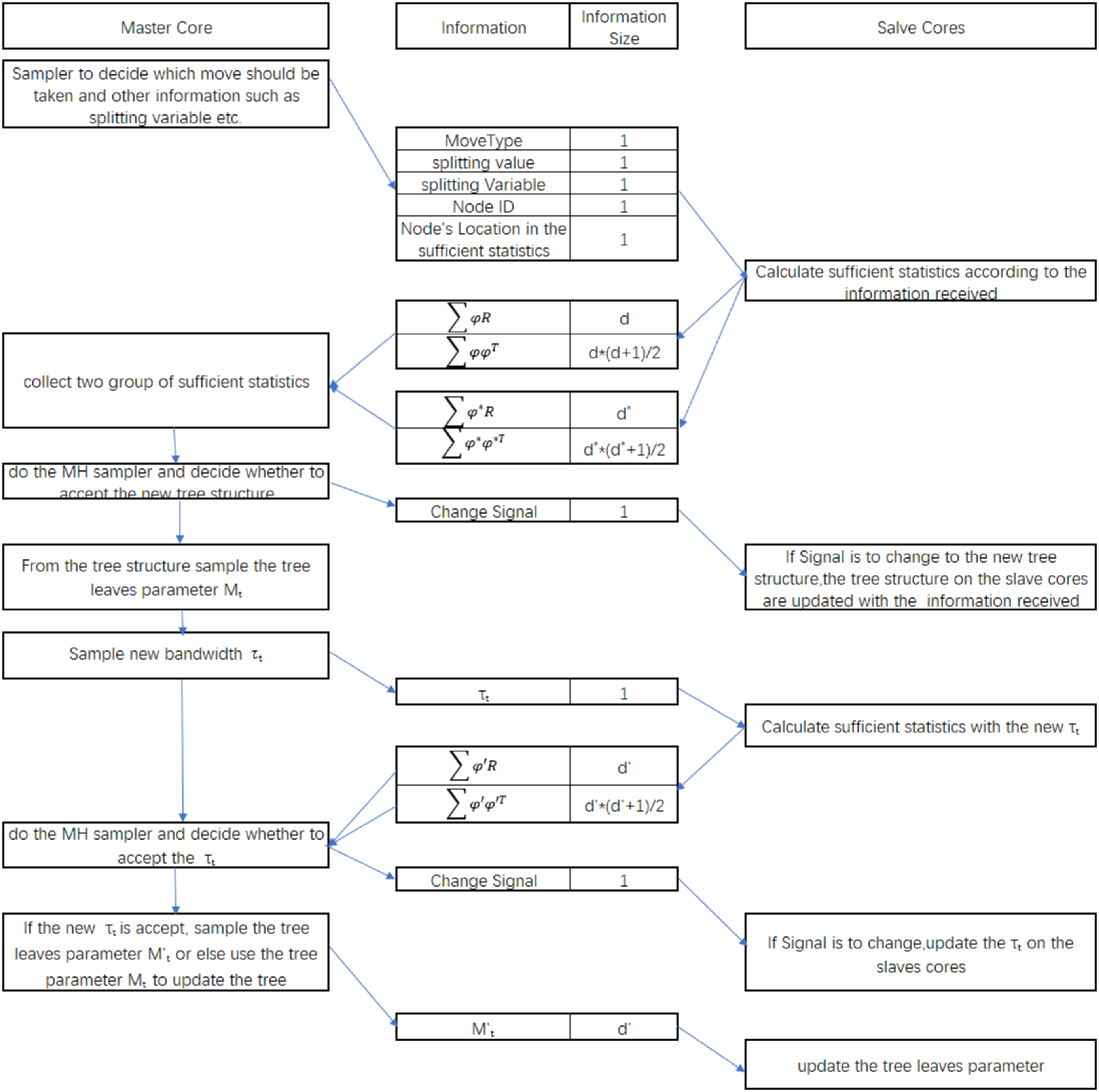}
	\end{center}
	\caption  {  \small{\it{ Summary of MCMC sampler step to update tree. The d,$d^{*}$,$d'$ denote the corresponding leaves count for the tree. }}}
	\label{FIG2}
\end{figure}
The main part for the MCMC sampler to update tree structure is illustrate in Figure \ref{FIG2} and we omit the update process for $(s,\sigma, \sigma_{\mu}, a)$.The communication path and information size for each step are listed.Compared to the parallel BART its communication cost increase a lot ,that is ,for about $O(d^{2})$ versus $(O(d))$ where $d$ denotes the count of leaves of the tree structure.Fortunately in the BART model we prevent the tree from growing too big.But this increasing communication cost  will lead to some problem especially in complicate communication environment.

We then show how we can use distributed SBART to speed up with additional processors.
Considering a large dataset $(x,y)$ with 100,000 records.Covariates x is consist of 220 covariables.Response variable is produced by
  \begin{eqnarray} \label{equ:fuction}
y_{i}=10sin(2\pi x_{i1}x_{i2} )+(x_{i3}-0.5)^{2} +x_{i4} +x_{i5} +\varepsilon_{i}
  \end{eqnarray}
x is i.i.d. and draws from a uniform distribution $U[0, 1]$.$\varepsilon_{i}$ is draw from i.i.d. normal distribution $N(0,1)$.We carry out 1,000 MCMC iterations with the first 500 discarded as burn-in.

\citet{pratola2014parallel} define the speedup of an algorithm by
\begin{eqnarray} \label{equ:spd1}
S(n, p)=\frac{T_{seq}}{T_{par}}
\end{eqnarray}
which is the ratio of the times taken to run two instances of the algorithms.
$T_{seq}$ stands for the  sequential algorithm’s executing time.$T_{par}$  stands for the  parallel algorithm’s executing time.
And the efficiency
\begin{eqnarray} \label{equ:spd2}
E(n, p)=\frac{S(n, p)}{p}
\end{eqnarray}
is the speedup normalized to the number of cores used in algorithm.n denote the total sample size and p denote the processors involved.
Here we can measure the efficiency relative to the speedup of 2 parallel cores.
\begin{eqnarray} \label{equ:spd3}
E(n, p)=\frac{2*T_{2}}{p*T_{p}}
\end{eqnarray}

We assign the job to different numbers of cores to see the effect.Table $\ref{tb2}$ shows the time required to carry out the MCMC draws as a function of the number of processors using this distributed SBART. Here total of 1,000 MCMC iterations were
carried out for each run. As expected, the running time decreases with the number of processors.The efficiency is kept in high level which stands for that this algorithm is scalable and we can speed up the algorithm by adding more workers.

\begin{table}
\begin{center}
\caption{Time to complete 1,000 MCMC iterations for a $100,000\times40$ dataset.The run time is in seconds.}
\label{tb2}
\begin{tabular}{ccc}
\hline
Cores & Run Time & Efficiency \\ \hline
2     & 1846   & 1          \\
4     & 986   & 0.94       \\
6     & 708         & 0.87       \\
8     & 543         & 0.85       \\
10    & 345        & 1.07       \\
20    & 180        & 1.02       \\
\hline
\end{tabular}

\end{center}
\end{table}

If speeding up the algorithm at big data situation is our main reason to choose this distributed algorithm.We can consider another option of embarrassingly parallel.We can easily do this by running copies of the algorithm at different worker. The downside is that the burn-in samples are not parallelized.For example we need to carry out 1000 iterations with the first 500 discarded as burn-in.Now we have 10 workers available.At each workers,we keep a copy of the whole dataset.For each worker we need to carry out 500 burn-in iterations and 50 iterations result independently with no communication with each other with different initial setting.Then combine all the result of the 10 workers together to form 500 valid iterations to poster estimation use.So from this point of view the embarrassingly parallel algorithm is not scalable and no matter how many workers are involved its efficiency is comparable to the distributed SBART with two workers in the previous example.That is one of the reason for us to develop distributed SBART.

\section{Data Experiment and Application}
Another improvement we made is that in the original SBART package they didn't provide solution for binary classification.Our package can solve binary classification problem and show its advantage versus BART.
\subsection{Data Experiment}

\begin{figure}

\begin{center}
    \includegraphics[width=0.7\textwidth]{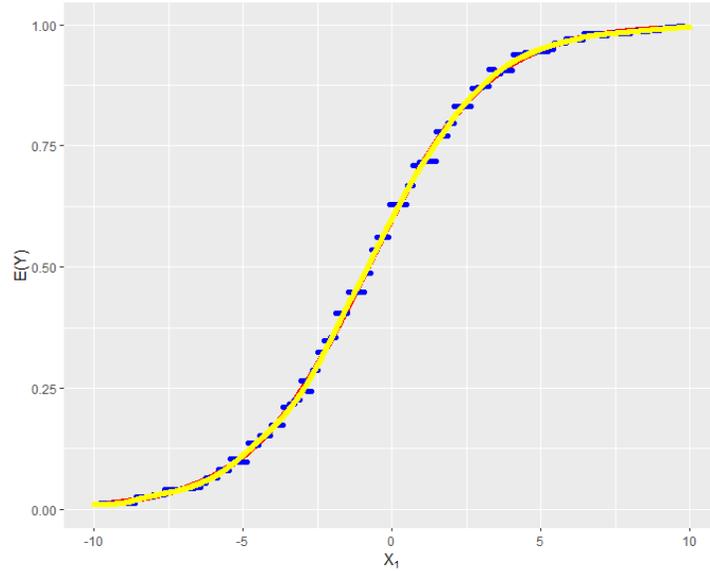}
\end{center}
\caption  {  \small{\it{ SBART VS BART.Red for true function,yellow  for SBART estimator and blue for BART estimator}}}
\label{FIG3}
\end{figure}

Considering dataset $(x,y)$ with 100,000 records.x consist of 10 covariables.These data are produced by
  \begin{eqnarray} \label{equ:binaryfuction}
E(y)=\frac{1}{e^{-(0.4+0.5X_{i1})}}
  \end{eqnarray}
x are i.i.d. draws from a uniform distribution $U[-10, 10]$.1,000 MCMC iterations are carried out with the first 500 discarded as burn-in.
We generate a test data set with 1,000 records.$X_{1}$ are grid points from -10 to 10 and other x are i.i.d. draws from a $U[-10, 10]$ distribution.

In figure $\ref{FIG3}$,we can find that the SBART nearly perfectly match with the true function.And the step function bias of BART can be easily captured which will do harm to the estimation process.From the example it is beneficial to apply SBART model for binary classification problem.

\subsection{Skin segmentation}
Skin segmentation is a real world data example from  UCI data sets\citep{Dua2019}.
For human object detection, skin segmentation is treated as a pre-processing step followed by the other
algorithms.  skin dataset is collected by randomly sampling B,G,R values from face images of various
age groups (young, middle, and old), race groups (white, black, and Asian) and genders obtained from FERET database
and PAL database.Total sample size is 245057 out of which 50859 is the skin samples and 194198 is non-skin samples.
We randomly select $80\%$ of the sample as training data set and leave the $20\%$ as test data set.We run the BART and distributed SBART
in 2000 iterations with the first 1000 discarded as burn-in.The BART algorithm costs about 845 seconds and the distributed SBART costs about 1896 seconds with 10 workers unless it will cost us more than 5 hours to finish this job with .

\begin{table}[]
\begin{center}
\caption{Confusing matrix for BART and SBART.}
\label{tb3}
\begin{tabular}{|c|c|c|c|c|}
\hline
\multirow{2}{*}{} & \multicolumn{2}{l|}{BART Prediction} & \multicolumn{2}{l|}{SBART Prediction} \\ \cline{2-5}
                  & Not Skin           & Skin            & Not Skin            & Skin            \\ \hline
Not Skin          & 38700              & 58             & 38690               & 68             \\ \hline
Skin              & 14                & 10239           & 7                  & 10246           \\ \hline
\end{tabular}
\end{center}
\end{table}

We get the confusion matrix as in Table $\ref{tb3}$ and find SBART perform a little better. That is ,the SBART reduced 7 samples of misclassification in the skin segment in exchange of 10 samples of misclassification in the not skin segment.Because the skin segment ratio is about $21\%$.So we call it a little improvement.

Test set performance for classification problem is measured by area under the Receiver Operating Characteristic curve(AUC), via the ROCR package of \cite{2015ROCR}. Larger AUC values indicate superior performance, with an AUC of 0.50 corresponding to the expected performance of a method that randomly orders observations by their predictions. A classifier’s
AUC value is the probability that it will rank a randomly chosen y = 1 example higher than a randomly chosen y = 0.

Test set performance for this example is measured by AUC.The SBART get the higher AUC result$(99.9934\%)$ than the BART algorithm $(99.9906\%)$.

\subsection{HIGGS Data Set}
\begin{figure}[htb]
	
	\begin{center}
		\includegraphics[width=0.6\textwidth]{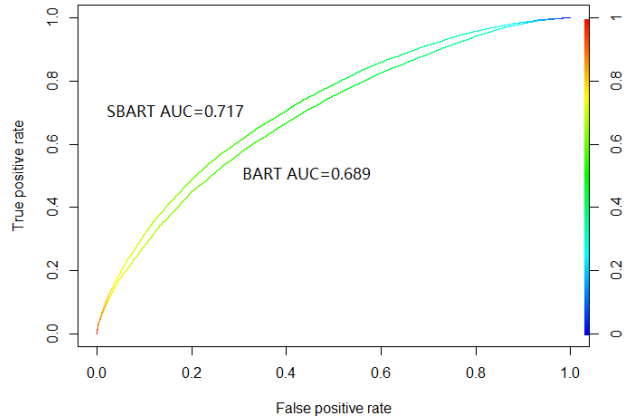}
	\end{center}
	\caption  {  \small{\it{ Performance comparison for BART and SBART model for Higgs data}}}
	\label{FIG4}
\end{figure}
Another real world data example from  UCI data sets\citep{Dua2019} is Higgs data set.The Higgs data has been produced using Monte Carlo simulations. The first 21 features (columns 2-22) are kinematic properties measured by the particle detectors in the accelerator. The last seven features are functions of the first 21 features; these are high-level features derived by physicists to help discriminate between the two classes.We use this data set for two purposes.One is to show its ability to handle big data.Another is to show the advantage of SBART compared to the BART algorithm in classification problem.So we sample 500,000 samples from the data set and apply SBART and BART algorithm to the data set.We randomly sample $90\%$ of the 500,000 as training data and the left $10\%$ as test data.We only analysis the low-level features.We run the BART and distributed SBART in 2000 iterations with the first 1000 discarded as burn-in.

From Figure $\ref{FIG4}$ we can see that the SBART model is superior to the BART model with higher AUC by overcoming the step-wise bias of the BART model.

\section{Conclusion and Looking Forward}

In this paper we have implemented a distributed SBART algorithm.The novelty of this paper lies in 3 folds
\begin{itemize}

	\item[$\bullet$] We accelerate the SBART algorithm and reduce the run time to less than $60\%$ of serial SBART and find a way to reduce the size of information without information lost. 
	\item[$\bullet$] SBART is expanded to distributed computation scenario under MPI framework.The distributed SBART can also work with observational datasets which may be too large to be stored in a single contiguous location.
	\item[$\bullet$] We expand SBART to binary classification problem which is not support in the original SBART package and 
demonstrate the advantage of SBART compared to BART in binary classification problem. 	
\end{itemize}

We also demonstrate the capabilities of the algorithm by data experiments and observed that the sampler's scalability.An example of 5 million observations of Higgs data demonstrated the algorithms ability to handle big data set.

A drawback for MPI is that it don't handle the fault.For example we are running distributed SBART under complicate situation and one of the worker break down,the whole process will be stopped and wait for dead node to wake up again. How to design a fault-tolerance algorithm  is an interesting topic.

For the Higgs data,the performance of SBART is comparable to boosted decision trees (0.73), shallow neural networks( 0.733) in the work of \cite{2014Searching},but there is a big gap between SBART and deep neural networks (0.88),what's the reason for such big gap and can we modify SBART and catch up with the deep neural networks is very attractive.


\newpage

\vskip20pt
\def\refhg{\hangindent=20pt\hangafter=1}
\def\refmark{\par\vskip 1mm\noindent\refhg}
\bibliographystyle{plainnat}

\bibliography{PSBART}

\end{document}